\documentclass[12pt]{article}
\usepackage{vmargin}
\setpapersize[portrait]{A4}            
\setmargins{0.9in}{0.5in}{6.5in}{9.2in}
           {0.3in}{0.3in}{0.3in}{0.3in}
\usepackage{fancyhdr}
\usepackage{latexsym}

\usepackage{epsfig}

\usepackage{pstcol}
\usepackage{pst-text}
\usepackage{pst-node}
\usepackage{pstricks}

\usepackage{shadow,epsf,amsthm,amssymb,amsmath}

\usepackage{enumerate}
\usepackage{graphicx}
\usepackage{hvdashln}                           
\usepackage{setspace}                           

\usepackage[matrix,frame,arrow]{xypic}
\input{Qcircuit}

\newtheorem{definitionenv}{Definition}
\newtheorem{lemmaenv}[definitionenv]{Lemma}
\newtheorem{theoremenv}[definitionenv]{Theorem}
\newtheorem{corollaryenv}[definitionenv]{Corollary}
\newtheorem{propositionenv}[definitionenv]{Proposition}
\newtheorem{conjectureenv}[definitionenv]{Conjecture}
\newtheorem{exampleenv}{Example}
\newtheorem{app-lemmaenv}[section]{Lemma}

\newenvironment{definition}{\begin{definitionenv}\rm}{\end{definitionenv}}
\newenvironment{lemma}{\begin{lemmaenv}\rm}{\end{lemmaenv}}
\newenvironment{theorem}{\begin{theoremenv}\rm}{\end{theoremenv}}
\newenvironment{corollary}{\begin{corollaryenv}\rm}{\end{corollaryenv}}
\newenvironment{example}{\begin{exampleenv}\rm}{\end{exampleenv}}
\newenvironment{proposition}{\begin{propositionenv}\rm}{\end{propositionenv}}
\newenvironment{conjecture}{\begin{conjectureenv}\rm}{\end{conjectureenv}}
\newenvironment{app-lemma}{\begin{app-lemmaenv}\rm}{\end{app-lemmaenv}}

\newcommand{\bd}{\begin{definition}}
\newcommand{\ed}{\end{definition}}
\newcommand{\bl}{\begin{lemma}}
\newcommand{\el}{\end{lemma}}
\newcommand{\elp}{\hspace*{\fill} $\Box$
                 \end{lemma}}
\newcommand{\bt}{\begin{theorem}}
\newcommand{\et}{\end{theorem}}
\newcommand{\etp}{\hspace*{\fill} $\Box$
                 \end{theorem}}
\newcommand{\bc}{\begin{corollary}}
\newcommand{\ec}{\end{corollary}}
\newcommand{\ecp}{\hspace*{\fill} $\Box$
                 \end{corollary}}
\newcommand{\bcj}{\begin{conjecture}}
\newcommand{\ecj}{\end{conjecture}}

\newcommand{\be}{\begin{example}}
\newcommand{\ee}{\end{example}}
\newcommand{\eep}{\hspace*{\fill} $\Box$
                 \end{example}}
\newcommand{\bp}{\begin{proposition}}
\newcommand{\ep}{\end{proposition}}
\newcommand{\epp}{
                 \end{proposition}}

\begin{document}

\title{
      Entanglement Increases the Error-Correcting Ability of Quantum Error-Correcting Codes
      }
\author{Ching-Yi Lai \ \ \mbox{and}\ \ Todd Brun
        \thanks{
        The authors are with the Communication
        Sciences Institute of the Ming Hsieh Electrical Engineering Department,
        University of Southern California,
        Los Angeles, CA 90089, USA \
        (E-mails:\ laiching@usc.edu and
        tbrun@usc.edu)
       }
       }
\date{\today}

\maketitle

\vspace{-1cm}
\begin{abstract}
If entanglement is available, the error-correcting ability of quantum codes can be increased.
We 
show how to optimize the minimum distance of an
entanglement-assisted quantum error-correcting (EAQEC) code,
obtained by adding  ebits to a standard quantum error-correcting
code, over different encoding operators. By this encoding
optimization procedure, we found  several new EAQEC codes,
including a family of $[[n,1,n;n-1]]$ EAQEC codes for  $n$ odd and
code parameters $[[7,1,5;2]]$, $[[7,1,5;3]]$, $[[9,1,7;4]]$,
$[[9,1,7;5]]$, which saturate the quantum singleton bound for
EAQEC codes.
A random search algorithm for  the encoding optimization procedure is also proposed.
\end{abstract}

\begin{flushleft}
\end{flushleft}


\section{Introduction} \label{sec:introduction}

Since Shor proposed the first quantum error-correcting code
\cite{Shor95}, the theory of quantum error correction has been
extensively developed. Today, quantum stabilizer
codes \cite{CRSS97,CRSS98,Got96,Got97,NC00} are  the most
widely-used class of quantum error-correcting codes.
One reason for this is that the CSS and CRSS code constructions \cite{CS96,Ste96,CRSS97,CRSS98}
allow classical self-orthogonal codes to be easily
transformed into quantum stabilizer codes.

Bowen  constructed the first entanglement-assisted quantum error-correcting (EAQEC) code from a three-qubit
bit-flip code with the help of two pairs of maximally-entangled states \cite{Bowen02}.
Bowen's code, which can correct an arbitrary one-qubit error, serves as an example that entanglement
increases the error-correcting ability of quantum
error-correcting codes.
Brun, Devetak and Hsieh showed that if shared entanglement between the
encoder and decoder is available,
classical linear quaternary (and binary)  codes that are not self-orthogonal can be transformed to EAQEC codes \cite{BDM06,BDM062}.

An $[n,k,d]$ classical linear quaternary code encodes $k$
 quaternary information digits into $n$ quaternary digits and can
correct up to $\lfloor \frac{d-1}{2}\rfloor$ quaternary digit
errors, where $d$ is called the minimum distance of the code.
Brun, Devetak and Hsieh showed that an $[n,k,d]$ classical linear quaternary code can be transformed to an $[[n, 2k-n+c, d; c]]$ EAQEC code
  that encodes $2k-n+c$ information qubits into $n$ qubits with the help of $c$ pairs of maximally-entangled states (ebits) for some $c$ \cite{BDM06}.
 This EAQEC code can correct up to $\lfloor \frac{d-1}{2}\rfloor$ qubit errors and has the same minimum distance $d$ as the classical code.
If entanglement is used, it boosts the rate of the code. However,
it has not been explored how entanglement can instead help
increase the minimum distance.
In addition, given parameters $n,k,c$, it is not clear how to  construct  an $[[n, k,d; c]]$ EAQEC code directly.
We will answer these questions in the paper.

An EAQEC code can be obtained from a standard QECC by changing
one or more ancilla qubits of the initial state with ebits.
We first discuss how adding the maximum number of ebits introduces symplectic partners of the generators of the stabilizer group, and
define a selection operator that determines the set of logical operators.
Then the minimum distance of the EAQEC codes can be optimized over distinct selection operators.
If we add fewer than the maximum ebits, we have the freedom to choose the set of generators of the stabilizer group,
 and the freedom to  replace different ancilla qubits with ebits.
EAQEC codes can be optimized over these choices.
These factors, together with the selection operator, can be represented in terms of unitary row operations.
 Applying this encoding optimization procedure to some standard quantum stabilizer codes, we construct several new EAQEC codes,
including a family of
$[[n,1,n;n-1]]$ EAQEC codes for  $n$ odd, and codes with parameters
$[[7,1,5;2]]$, $[[7,1,5;3]]$, $[[9,1,7;4]]$, $[[9,1,7;5]]$, which
saturate the quantum singleton bound for EAQEC codes,
and  are not equivalent to any
standard quantum stabilizer code. (We say that an $[[n,k,d;c]]$
EAQEC code is not equivalent to any standard quantum stabilizer
code if there is no standard $[[n+c,k,d]]$ quantum code.)
We also compare the EAQEC codes obtained by the encoding optimization procedure 
with  the EAQEC codes obtained by the construction of \cite{BDM06}. 

Although the encoding optimization procedure seems to be a promising method to construct EAQEC codes
with high minimum distance $d$ for given parameters $n,k,c$,
it will be shown that the complexity of the encoding optimization procedure increases exponentially with $n+k$,
which implies that it is impossible to fully optimize the minimum distance for high $n+k$.
Hence we develop a random search algorithm, which achieves a suboptimal result efficiently.
On the other hand, we perform a different computer search for EAQEC codes
that have  a circulant check matrix 
and find several EAQEC codes that achieve the quantum singleton bound and are not equivalent to any standard quantum stabilizer code.
This circulant construction of EAQEC codes also provides evidence that entanglement helps increase
the error-correcting ability of the quantum error-correcting codes.

This paper is organized as follows.
Basics of stabilizer codes and EAQEC codes are introduced in Section \ref{sec:preliminary}.
We discuss the encoding optimization procedure by first considering the case
of maximal entanglement and the selection operator in Section \ref{Sec:Encoding_Circuit},
including the construction of the family of $[[n,1,n;n-1]]$ EAQEC codes for  $n$ odd. 
We then generalize to adding arbitrary amounts of entanglement.
The effect of unitary row operations are discussed in Section \ref{Sec:Row_operation}, which completes the encoding optimization procedure.
The results of applying the encoding optimization procedure to some standard quantum  stabilizer codes are provided in Section \ref{sec:result}.
Then we propose the random search algorithm for EAQEC codes in Section \ref{sec:random_search}, together with several examples.
The circulant construction of EAQEC codes is described in Section \ref{sec:circulant_construction}, followed by the discussion section.



\section{Preliminaries} \label{sec:preliminary}
\subsection{Stabilizer Codes}
Suppose $\mathcal{S}$ is an abelian subgroup of Pauli group $\mathcal{G}_n$ that does not include $ -I$,
with a  set of $r\equiv n-k$ independent generators $\{ g_1,g_2,\cdots,g_r\}$.
An $[[n,k,d]]$ quantum stabilizer code $\mathcal{C(S)}$ corresponding to the stabilizer group $\mathcal{S}$
is the $2^k$-dimensional subspace of the $n-$qubit state space fixed by  $\mathcal{S}$.
The minimum distance $d$ is the minimum weight of an element in $\mathcal{N}(\mathcal{S})-\mathcal{S}$,
where $\mathcal{N}(\mathcal{S})$ is the normalizer group of $\mathcal{S}$.
An element $g=i^{m} M_1\otimes M_2\otimes \ldots \otimes M_n$ in
$\mathcal{G}_n,$ where $M_i\in \{I, X, Y, Z\}$ and $m \in \{0,1,2,3\}$, can be expressed
as $g=i^{m'} X_{\alpha}Z_{\beta}$ with $\alpha, \beta$ two binary $n$-tuples and $m' \in \{0,1,2,3\}$.
In this expression, if $M_j=I,$ $X$, $Z$, or $Y$, then the $j-$th bits of $\alpha$ and $\beta$  are
$(\alpha_j, \beta_j)=(0,0)$, $(1,0)$, $(0,1)$, or $(1,1)$, respectively, and $m'\equiv m+l \ (\mbox{mod}\ 4)$,
where $l$ is the number of $M_j$'s equal to $Y$.
We define a homomorphism $\varphi : \ \mathcal{G}_n \mapsto \mathbb{Z}^{2n}_2 $ by 
\[
    \varphi(i^m X_{\alpha}Z_{\beta})=(\alpha,  \beta),
\]
and define a symplectic inner product $\odot$ between two elements $(\alpha_1,\beta_1)$ and $(\alpha_2,\beta_2)$ in  $\mathbb{Z}_2^{2n}$ by
\begin{align*} 
    (\alpha_1,\beta_1)\odot (\alpha_2,\beta_2)\triangleq \alpha_1\cdot \beta_2+ \beta_1 \cdot \alpha_2,
\end{align*}
where $\cdot$ is the usual inner product in $\mathbb{Z}_2^{n}$.
Thus, two elements $g,h$ in $\mathcal{G}_n$ commute if and
only if the symplectic inner product $\varphi(g) \odot \varphi(h)$  is zero.
Then a check matrix $H$  corresponding to the stabilizer $\mathcal{S}$ is defined as a binary $r\times 2n$ matrix
such that the the $i-$th row vector of $H$ is $\varphi(g_i)$.
For convenience,  $H$ is denoted by
\begin{align*}
[ H_X | H_Z],
\end{align*}
where $H_X$ and $ H_Z$ are two $r\times n$ binary matrices.
The check matrix $H$ must satisfy the following
commutative condition,
\begin{align*}
    H \Lambda_{2n} H^T = H_X H_Z^T+H_ZH_X^T=O_{r\times r}, \label{eqn:commutative}
\end{align*}
where $\Lambda_{2n} =
\begin{bmatrix}
O_{n\times n} & I_{n\times n}\\
I_{n\times n} & O_{n\times n}
\end{bmatrix}$, $ O_{i\times j}$ is an $i\times j$ zero matrix, and $ I_{r\times r}$ is an $r$-dimensional identity matrix.
The $i$-th column of $H_X$ is the error syndrome of the error operator $Z_i$ and the  $j$-th column of $H_Z$
is the error syndrome of the error operator $X_j$.
The error syndrome of $Y_l$ is the sum of the $l$-th column of $H_X$ and the $l$-th column of $H_Z$.
In general, the error syndrome of  $g=i^{m} M_1\otimes M_2\otimes \ldots \otimes M_n$ in
$\mathcal{G}_n,$ where $M_i\in \{I, X, Y, Z\}$, is the linear combination of the error syndromes of each factor $M_i$.
For a code with minimum distance $d$, if the error syndromes of error operators of weight smaller than or equal to
$\lfloor \frac{d-1}{2}\rfloor$ are distinct,
we call that code nondegenerate.
Otherwise, it is degenerate.

The encoding procedure is described as follows. Consider the initial $n-$qubit state
 \[\ket{\psi}=\underbrace{\ket{0}\ket{0}\cdots\ket{0}}_{r=n-k}\ket{\phi} ,\]
where  there are $r=n-k$ ancilla qubits $\ket{0}$'s and an arbitrary $k-$qubit state $\ket{\phi}$.
A set of generators of the stabilizer
group of this class of states is
\begin{align*}
\begin{split}
Z_1=&ZIIII\cdots I\\
Z_2=&IZIII\cdots I\\
&\vdots\\
Z_r=&I\cdots IZ \underbrace{I\cdots I}_{k}
\end{split}
\end{align*}
with a  check matrix  
\begin{align} \label{initial_stabilizer_check_matrix}
\left[
\begin{array}{c|cc}
O_{r\times n} & I_{r\times r} & O_{r\times (n-r)}\\
\end{array}
 \right].
\end{align}
The logical operators on $\ket{\psi}$ are
\[
Z_{r+1}, \cdots, Z_{n},
\]
and
\[
X_{r+1}, \cdots, X_{n}.
\]
If $U_E$ is a  unitary operator such that
$\{U_EZ_1U_E^{\dag}, \cdots, U_EZ_rU_E^{\dag} \}$ is a set of generators of the stabilizer group $\mathcal{S}$,
then $U_E$ is an encoding  operation of $\mathcal{C(S)}$, and the encoded state $U_E \ket{\psi}$ is fixed by the stabilizer group $\mathcal{S}$.
In particular, we can choose
\begin{align*}
 g_i=U_EZ_iU_E^{\dag}
\end{align*}
for $i=1, \cdots, r$. 
The logical operators on $U_E\ket{\psi}$ are
\[
\bar{Z}_1= U_EZ_{r+1}U_E^{\dag}, \cdots, \bar{Z}_k= U_EZ_{n}U_E^{\dag},
\]
\[
\bar{X}_1= U_EX_{r+1}U_E^{\dag}, \cdots, \bar{X}_k= U_EX_nU_E^{\dag}.
\]
Note that the logical operators commute with the stabilizers, and the normalizer group of $\mathcal{S}$ is
\[
\mathcal{N(S)}= \langle g_1,g_2,\cdots, g_r, \bar{Z}_1,\bar{Z}_2,\cdots, \bar{Z}_k, \bar{X}_1,\bar{X}_2,\cdots, \bar{X}_k  \rangle
\]
with dimension $2n-r=r+2k.$

Given a check matrix $H$ of a stabilizer group, Wilde gave an algorithm \cite{wilde-2008} to
find an encoding circuit for this quantum stabilizer code.
This algorithm applies a series of CNOT gates, Hadamard gates, Phase gates, SWAP gates, and row operations
to the check matrix $H$  such that $H$ takes the form (\ref{initial_stabilizer_check_matrix}).
This process is like performing  Gaussian elimination on a matrix,
 but using  CNOT gates, Hadamard gates, Phase gates, and SWAP gates, in addition to the elementary row operations of Gaussian elimination.
 There are two types of elementary row operations over the binary field:
 adding one row  to another, which corresponds to  multiplying an operator by another; and
 exchanging two rows, which corresponds to relabeling two generators.
Performing these row operations does not change the row space and hence the error-correcting ability of the codes.
The effects of these gate operations  on the entries in the check matrix are as follows:
\begin{enumerate}
\item A CNOT gate from qubit $i$ to qubit $j$ adds column $i$ to
column $j$ in $H_X$ and adds column $j$ to column $i$ in $H_Z$.
\item A Hadamard gate on qubit $i$ swaps column $i$ in $H_Z$  with
column $i$ in $H_X$.
\item A Phase gate on qubit
$i$ adds column $i$ in  $H_X$ to column $i$ in $H_Z$.
\item Three
CNOT gates implement a SWAP gate. The effect of a
SWAP gate on qubits $i$ and $j$ is to swap columns $i$ and $j$ in both
$H_X$ and $H_Z$.
 use the check matrices such that no Row operations in the encoding algorithm.
\end{enumerate}
The series of operations used in the algorithm serve as a unitary operation $U_E^{\dag}$ such that  $U_E^{\dag}g_i U_E=Z_i$,
and hence  the inverse operator $U_E$ is a desired encoding operation.

Note that $U_E$ is not unique. The encoding algorithm generates, in general,
different $U_E$'s for different sets of generators of the same stabilizer group;
and performing the steps in the encoding algorithm in a different order will, in general,
generate a different $U_E$ even for the same set of generators.


\subsection{Entanglement-Assisted Quantum Error-Correcting Codes}
Brun, Devetak and Hsieh proposed a theory of quantum stabilizer codes when shared entanglement between
the encoder (Alice) and  decoder (Bob) is available in \cite{BDM06}. 
Suppose that  Alice and Bob share $c$ pairs of maximally entangled
states.
Assume further that Bob's halves of the $c$ ebits  are not subject to any error.
Now if an arbitrary subset $T=\{t_1,\cdots, t_c\}$ of the $r=n-k$
ancilla qubits  in $\ket{\psi}$ are replaced  by these $c$ pairs
of maximally-entangled states $\ket{\Phi_+}^{AB}$'s,
then the $(n+c)$-qubit initial state is
\begin{align*} \label{unencoded_EA_state}
\ket{\psi}_{EA}=\bigotimes_{i=1}^r\ket{\eta_i}\otimes
\ket{\phi} ,
\end{align*}
where\[\ket{\eta_i}=\left\{%
\begin{array}{ll}
    \ket{0}, & \hbox{if $i\notin T $;} \\
    \ket{\Phi_+}^{AB}, & \hbox{if $i\in T$.} \\
\end{array}%
\right.    \]
For convenience, the qubits on Alice's side will be
numbered 1 to $n$ and the qubits on Bob's side will be numbered 1
to $c$. Hence the $t_i$-th qubit of Alice and the $i$-th qubit of
Bob form a pair of maximally-entangled state.
Then a set of independent generators of a stabilizer group of $\ket{\psi}_{EA}$ is 
\begin{align}
\begin{split}
 \left\{ %
\begin{array}{ll}
    Z_i^A\otimes I^B,& \ \hbox{if $i\notin T$;} \\ \label{unencoded_EA_state stabilizer}
    Z_i^A\otimes Z_j^B,& \ \hbox{if $i=t_j \in T$; } \\
\end{array}%
\right.
&\mbox{for $i=1,\cdots, r$,}\\
   X_{t_j}^A \otimes X_j^B, \ \ \ \ \ \ \ \ \ \ \ \ \ \ \ \ \ \  \ \ \ \  & \mbox{for $j=1,\cdots, c$.}\\
\end{split}
\end{align}
Note that the operators on the left and right of the tensor product $\otimes$
are applied to Alice's qubits and  Bob's qubits, respectively, and the superscripts $A$ and $B$ will be omitted  throughout the rest of this article.
The logical operators on $\ket{\psi}_{EA}$ are $Z_{r+1}\otimes I$, $\cdots$, $Z_{n}\otimes I,$ and $X_{r+1}\otimes I$, $\cdots,$ $X_{n}\otimes I$.
Now consider the operators on Alice's qubits.
It can be observed that
\begin{align}
&[Z_i,Z_j]=0, \mbox{ for $0\leq i, j \leq r$},  \label{def:symplectic pairs 1} \\ %
&[X_{t_i},X_{t_j}]=0, \mbox{ for $0\leq i, j \leq c$},  \label{def:symplectic pairs 2} \\ %
&\{Z_{t_i},X_{t_i}\}=0, \mbox{  for $0\leq i \leq c$}, \label{def:symplectic pairs 3} \\
&[Z_i,X_{t_j}]=0, \mbox{ for $i\neq t_j$}, \label{def:symplectic pairs 4}
\end{align}
where  $[g,h]=gh-hg$ and $\{g,h\}=gh+hg$,
or
\begin{align}
&\varphi(Z_i) \odot \varphi(Z_j)=0, \mbox{ for $0\leq i, j \leq r$},  \label{def:symplectic pairs 5} \\ %
&\varphi(X_{t_i}) \odot  \varphi(X_{t_j})=0, \mbox{ for $0\leq i, j \leq c$},  \label{def:symplectic pairs 6} \\ %
&\varphi(Z_{t_i})\odot  \varphi(X_{t_i})=1, \mbox{  for $0\leq i \leq c$}, \label{def:symplectic pairs 7} \\
&\varphi(Z_i) \odot  \varphi(X_{t_j})=0, \mbox{ for $i\neq t_j$}. \label{def:symplectic pairs 8}
\end{align}
If a set of $(r+c)$ operators satisfy equations (\ref{def:symplectic pairs 1})-(\ref{def:symplectic pairs 4}) or
equations (\ref{def:symplectic pairs 5})-(\ref{def:symplectic pairs 8}),
we say that the two operators in (\ref{def:symplectic pairs 3}) or the two vectors in (\ref{def:symplectic pairs 7})
form a \emph{symplectic pair}, and they are \emph{symplectic partners} of each other.
Hence $Z_{t_i} $ and $X_{t_i}$ form a symplectic pair. 

An encoding operation $U_E$ is applied on Alice's $n$ qubits,
while no operation is performed on Bob's $c$ qubits. A set of
generators of a stabilizer group $\mathcal{S}$ of the encoded
state  $(U_E\otimes I)\ket{\psi}_{EA}$ is $\{g_1, \cdots, g_r, h_1, \cdots,
h_c\}$, where
\begin{align*}
 g_i=\left\{%
\begin{array}{ll}
    &U_EZ_iU_E^{\dag}\otimes I, \ \hbox{if $i\notin T$;} \\
    &U_EZ_iU_E^{\dag}\otimes Z_j, \ \hbox{if $i=t_j \in T$ } \\
\end{array}%
\right.
\end{align*}
and
\begin{align*}
h_j=
    U_EX_{t_j}U_E^{\dag}\otimes X_j.
\end{align*}
The logical operators on  $(U_E\otimes I)\ket{\psi}_{EA}$ are
\[
\bar{Z}_1= U_EZ_{r+1}U_E^{\dag}\otimes I, \cdots, \bar{Z}_k= U_EZ_{n}U_E^{\dag}\otimes I,
\]
and
\[
\bar{X}_1= U_EX_{r+1}U_E^{\dag}\otimes I, \cdots, \bar{X}_k= U_EX_nU_E^{\dag}\otimes I.
\]

The $2^k-$dimensional subspace of the $(n+c)-$qubit state space fixed by the stabilizer group $\mathcal{S}$
with a set of generators $\{g_1, \cdots, g_r, h_1, \cdots, h_c\}$
is called an entanglement-assisted quantum error-correcting  (EAQEC) code  with parameters $[[n,k,d;c]]$ for some minimum distance $d$.
With the help of $c$ pairs of maximally-entangled states, the stabilizer group of an $[[n,k,d;c]]$ EAQEC code
has $c$ more generators than that of an $[[n,k,d]]$ standard quantum stabilizer code.
The operators on Bob's side have no direct effect on the error-correcting ability of the EAQEC code if
we assume that the $c$ qubits of Bob suffer  no error.
These operators serve to make the stabilizer abelian, so that the generators can be measured simultaneously.
For convenience, we denote
\[
    g_i'= U_EZ_iU_E^{\dag},
\]
and
\[
 h_j'= U_EX_{t_j}U_E^{\dag},
\]
 and the $g_{i}'$s and $h_j'$s will be called the \emph{simplified generators} of the stabilizer group.
Similarly, we denote  $\bar{Z}'_i= U_EZ_{r+i}U_E^{\dag}, \bar{X}'_j= U_EX_{r+j}U_E^{\dag}$.
It is obvious that $\{g_1', \cdots, g_r',h_1', \cdots, h_c'\}$ satisfy the commutative
relations (\ref{def:symplectic pairs 1})-(\ref{def:symplectic pairs 4}), 
and $g_{t_i}'$ and $h_i'$ are a symplectic pair.
Let $\mathcal{S}'=\langle g_1', \cdots, g_r',h_1', \cdots, h_c'\rangle $, and $\mathcal{S}'_I=\langle g_j: j\notin T \rangle .$
Hence the normalizer group of $\mathcal{S}'$ is
\[
\mathcal{N(S')}= \langle g_{i}:i\notin T,  \bar{Z}_1',\cdots, \bar{Z}_k', \bar{X}_1',\cdots, \bar{X}_k'  \rangle
\]
with dimension $2n-(r+c)=2k+r-c.$
The minimum distance $d$ of the EAQEC code defined by $\mathcal{S}$  is the
minimum weight of an element in
$\mathcal{N(S')}-\mathcal{S}'_I$. 
In particular, when $c=r$, we have $\mathcal{S}'_I=\varnothing$ and
\[
\mathcal{N(S')}= \langle \bar{Z}_1',\cdots, \bar{Z}_k', \bar{X}_1',\cdots, \bar{X}_k'  \rangle.
\]
An $[[n,k,d;c]]$ EAQEC code must satisfy the
quantum singleton bound for EAQEC codes \cite{BDM06}:
\begin{align}
n+c-k\geq 2(d-1). \label{eq:singleton_bound}
\end{align}

We define a simplified check matrix $H'$ 
as a binary $(r+c)\times 2n$ matrix such that the $r+c$ row vectors of $H'$ are $\varphi(g_i')$ for $i=1, \cdots, r$ and $\varphi(h_{j}')$ for $j=1, \cdots, c$.
For simplicity, we usually order the generators $g_i'$ and $h_j'$ so that $\varphi(g_i')$ is the $i-$th row vector of $H'$  for $i=1, \cdots, r$,
 $\varphi(h_{j}')$ is the $(j+r)-$th row vector of $H'$ for $j=1, \cdots, c$, and the $j$-th and $(j+r)-$th row vectors are a symplectic pair.
$H'$ must satisfy  the commutation relations (\ref{def:symplectic pairs 5})-(\ref{def:symplectic pairs 8}), and in the case $c=r$,
\begin{align} \label{eq:commutation_relations_H}
H'\Lambda_{2n} H'^T=
\left[
\begin{array}{c|c}
O_{r \times r} & I_{r \times r} \\
\hline
I_{r\times r} &O_{r\times r}\\
\end{array}
 \right].
\end{align}
For example,  the simplified check matrix corresponding to the set of generators (\ref{unencoded_EA_state stabilizer}) of a stabilizer group of the initial state $\ket{\psi}_{EA}$  is
\begin{align} \label{eq:initial_stabilizer_check_matrix_EAQECC}
\left[
\begin{array}{c|c}
O_{r\times n} & I_{r\times r} \ O_{r\times (n-r)}\\
\hline
I_{r\times r} \ O_{r\times (n-r)} &O_{r\times n}\\
\end{array}
 \right].
\end{align}
Conversely, an $(r+c)\times 2n$ binary matrix $\tilde{H}$, serving as a simplified check matrix, can define a stabilizer group and hence an EAQEC code.
The number of ebits required to construct an EAQEC code
\cite{WB-2008-77} is 
\begin{align} \label{eq:number of ebits}
c=\frac{1}{2}\mbox{rank}(\tilde{H}\Lambda \tilde{H}^T).
\end{align}
Like a check matrix of a standard quantum error-correcting code,
the simplified check matrix $H'$ can be used to determine the minimum distance of  nondegenerate EAQEC codes. 
Note that Wilde's encoding circuit algorithm \cite{wilde-2008} can also be applied to a simplified check matrix to find an encoding unitary operator of the EAQEC code, just as for a standard stabilizer code.

Similarly, we define a simplified logical matrix $L'$
corresponding to the logical operators by
putting $\varphi(\bar{Z}_i')$ to be the $i-$th row vector of $L'$  for $i=1, \cdots, k$,
and   $\varphi(\bar{X}_{j}')$ to be the $(j+k)-$th row vector of $L'$ for $j=1, \cdots, k.$
Since the logical operators commute with $\{ g_1', \cdots, g_r',h_1', \cdots, h_r'\} $, we have
\begin{align}
H'\Lambda_{2n} L'^T = O. \label{eq:H Lambda GL}
\end{align}
Since the logical operators  satisfy the commutative relations (\ref{def:symplectic pairs 1})-(\ref{def:symplectic pairs 4}), we have
\begin{align*}
L'\Lambda_{2n} L'^T=
\left[
\begin{array}{c|c}
O_{k \times k} & I_{k \times k} \\
\hline
I_{k\times k} &O_{k\times k}\\
\end{array}
 \right].
\end{align*}
For example,  the simplified logical matrix corresponding to the initial state $\ket{\psi}_{EA}$  is
\begin{align} \label{eq:initial_logical_matrix_EAQECC}
\left[
\begin{array}{c|c}
O_{k\times n} & \ O_{k\times r}\ I_{k\times k} \\
\hline
\ O_{k\times r}\ I_{k\times k}  &O_{k\times n}\\
\end{array}
 \right].
\end{align}

\section{The Encoding Optimization Procedure for EAQECCs} 

A standard $[[n,2k-n,d]]$ quantum stabilizer code, obtained by the CSS construction \cite{CS96,Ste96} from an $[n,k,d]$ classical linear self-orthogonal code,
has the same minimum distance as the classical code used.
On the other hand, an $[[n,2k+c-n,d;c]]$ EAQEC code can be constructed from an $[n,k,d]$ classical linear quaternary  code  by the construction of \cite{BDM06}, and $c$ is determined by (\ref{eq:number of ebits}).
It seems that only the number of information qubits is increased by introducing ebits.
However, with the help of entanglement it is possible to define more distinct error syndromes for a given codeword size, and hence the set of correctable error operators might be larger.
We would like to construct EAQEC codes with a higher minimum distance instead of higher rate.

One way to construct an EAQEC code is to start with a standard QECC and move $c$ of the qubits from Alice's side to Bob's side.  So long as $c\le d/2$, the resulting code can be encoded by a unitary operator on Alice's side, given $c$ ebits of initial shared entanglement between Alice and Bob.  While such codes can be interesting and useful, they are not the subject of interest for this paper; because such codes retain an ability to correct errors on Bob's qubits, they are in a sense not making full use of the fact that Bob's halves of the ebits are noise-free.  They therefore may not have the maximum error correcting power on Alice's qubits for the given parameters $n$, $k$ and $c$.  We are interested in EAQEC codes that can do better than any standard code in this sense.

To make this idea precise, we say that an $[[n,k,d;c]]$ EAQEC code is not equivalent to any standard quantum stabilizer code if there is no standard $[[n+c,k,d]]$ quantum code.  If there exists a standard $[[n+c,k,d]]$ quantum code, then we may not be achieving the maximum boost to our error correcting power from the $c$ ebits of shared entanglement.   We expect  added entanglement in general to increase the error-correcting ability of a quantum error-correcting code such that the EAQEC code is not equivalent to any standard quantum stabilizer code, and indeed this turns out to be possible by our encoding optimization procedure.  (Note that this is not {\it always} possible---the smallest examples of the $[[3,1,3;2]]$ and $[[4,1,3;1]]$ codes are both equivalent to the standard $[[5,1,3]]$ QECC, and this is the best that can be done.)


We now consider how added entanglement affects standard quantum stabilizer codes.
Suppose $\{g_1',g_2',\cdots, g_r'\}$ forms a set of independent generators of a stabilizer group $\mathcal{S}$ for an $[[n,k,d]]$ quantum stabilizer code $\mathcal{C(S)}$,
and $U_E$ is the encoding unitary operator obtained by Wilde's encoding circuit algorithm \cite{wilde-2008}, such that $U_E Z_1U_E^{\dag}= g_1',\cdots, U_EZ_rU_E^{\dag}=g_r'$.
The logical operators are  $\bar{Z}'_1= U_EZ_{r+1}U_E^{\dag},$ $ \cdots,$ $\bar{Z}'_k= U_EZ_{n}U_E^{\dag},$ $\bar{X}'_1= U_EX_{r+1}U_E^{\dag},$ $\cdots, $ $\bar{X}'_k= U_EX_{n}U_E^{\dag}$.
Suppose entanglement between the encoder and decoder is available, and a set $T$ of $c$ ancilla qubits are replaced by ebits.
An EAQEC code is obtained.
This introduces $c$ simplified generators $h_j'=U_E X_{t_j}U_E^{\dag}$'s, $t_j\in T$, $j=1, \cdots, c$,  to the generating set of the stabilizer group $\mathcal{S}$.
As we will examine in detail below, the encoding unitary operator is not uniquely defined.
The EAQEC code defined by $S'=\langle g_1', \cdots, ,g_r',h_1', \cdots, ,h_c'\rangle$ may gain higher error-correcting ability by modifying the encoding operator.

We first discuss the case $c=r$, where the generators $h_i'$'s are
symplectic partners of $g_i'$'s, respectively. We will treat  the
case $c<r$ later, by optimizing the choice of $c$ linearly independent generators from the group $\langle h_1', \cdots, ,h_r'\rangle$.

\subsection{Selecting Symplectic Partners and Logical Operators}  \label{Sec:Encoding_Circuit}
Since the symplectic partners of $g_1', \cdots, g_r'$ are not unique, we now explain how to select these partners  such that the minimum distance of the EAQEC codes is higher than the code without entanglement.
Suppose $W$ is a unitary  Clifford operator that commutes with $Z_1, \cdots, Z_r$ such that after the operation of $W$,
the simplified check matrix of the initial state (\ref{eq:initial_stabilizer_check_matrix_EAQECC})
becomes
\begin{align} \label{eq:reduced_encoding_check_matrix}
\left[
\begin{array}{c|cc}
 O_{r\times n}  & I_{r\times r} & O_{r\times(n-r)}\\
\hline
 I_{r\times r} \ A &  C & B\\
\end{array}
\right],
\end{align}
where 
$A$ and $B$ are two $r\times(n-r)$ binary matrices, and $C$ is an $r\times r$ binary matrix.
The simplified check matrix satisfies the commutation relations (\ref{def:symplectic pairs 5})-(\ref{def:symplectic pairs 8}) with
\begin{align}
C^{T}+AB^{T}+C+BA^{T}=O. \label{eq:ABC}
\end{align}
In addition, it can be checked that
the simplified logical matrix is of the form
\begin{align*}
\left[
\begin{array}{c|cc}
 O_{k\times n}  & A^T & I_{k\times k}\\
\hline
O_{k\times n} \  I_{k\times k}  &  B^T & O_{k\times k}\\
\end{array}
\right]
\end{align*}
after Gaussian elimination such that (\ref{eq:H Lambda GL}) and (\ref{eq:initial_logical_matrix_EAQECC}) hold.
Since
\[
(U_E W) Z_i (U_EW)^{\dag}= U_E Z_i U_E^{\dag}=g_i'
\]
for $i=1, \cdots, r$,
$U_E W$ is also an encoding operator  of the quantum stabilizer code $\mathcal{C(S)}$.
However, the  symplectic partners of $g_i'$'s,
\[
U_E (WX_i W^{\dag})U_E^{\dag}
\]
may differ from $U_E X_i U_E^{\dag}$ for $i=1,$ $\cdots,$ $r$ and the logical operators
\[
U_E (WX_i W^{\dag})U_E^{\dag}, \
U_E (WZ_j W^{\dag})U_E^{\dag},
\]
for $i,j=r+1,$ $\cdots,$ $n$  are different.
Choosing  a set of matrices $A$, $B$, $C$ such that $C^{T}+AB^{T}+C+BA^{T}=O$
determines a unitary operator $W$ by the encoding circuit algorithm, which determines a set of symplectic partners of $g_1', \cdots, g_r'$ and a different set of logical operators.
Thus we call $W$ the \emph{selection operator} for EAQEC codes.
The minimum distance of the EAQEC code can be optimized by examining  each distinct encoding operator $U_E W$.
Note that the simplified logical matrix is not affected by the matrix $C$.
Therefore, there are 
 $2^{2rk}$ distinct sets of logical operators.
\bl
Given matrices $A$ and $B$,  then the matrix $C$, satisfying (\ref{eq:ABC}), is of the form
\begin{align*}
C= BA^T+M,
\end{align*}
where $M$ is a symmetric matrix.
\el
\begin{proof}
Suppose $C'=BA^T+M+M'$, where $M'$ is not a symmetric matrix, satisfies (\ref{eq:ABC}). Then
\begin{align*}
O=AB^T+BA^T+C'+(C')^T
= M'+(M')^T,
\end{align*}
which implies that $M'$ is the zero matrix.
\end{proof}
%
%
%
%
We construct an EAQEC code that achieves the quantum singleton bound by applying this procedure to a standard quantum stabilizer code in the following example.
\be  \label{example:5_1_5_4}
A check matrix of the standard $[[5,1,1]]$ 5-qubit bit flip code is
\begin{align*} 
\left[
\begin{array}{c|c}
00000&11000\\
00000&01100\\
00000&00110\\
00000&00011\\
\end{array}
 \right].
\end{align*}
Applying the encoding circuit algorithm to this check matrix, we obtain an encoding operator $U_E$.
In particular, if $C=O$ in (\ref{eq:ABC}), then
\begin{align*}
A B^T + B A^T =O. \label{eq:AB_c=0}
\end{align*}
When $k=1,$ 
$A B^T + B A^T =O$
holds if and only if $A=B$ or at least one of $A$ and $B$ is the zero vector.
Let $W$ be the selection operator determined by the encoding circuit algorithm with
$\displaystyle A=\begin{bmatrix}0&0&0&0\end{bmatrix}^{T}$ and  $\displaystyle B=\begin{bmatrix}1&0&1&0\end{bmatrix}^{T}$.
Then  the encoding operator  $U_EW$ generates a $[[5,1,5;4]]$ EAQEC code with a simplified check matrix
\begin{align*}
\left[
\begin{array}{c|c}
00000&11000\\
00000&01100\\
00000&00110\\
00000&00011\\
\hline
01111 & 00000\\
11000 & 00000\\
00011 & 00000\\
11110 & 00000\\
\end{array}
 \right]
\end{align*}
and a simplified logical matrix
\begin{align*}
\left[
\begin{array}{c|c}
11111 & 00000\\
\hline
00000&11111\\
\end{array}
 \right].
\end{align*}
With the help of $4$ ebits, the minimum distance is increased from $1$ to $5$.
The quantum singleton bound (\ref{eq:singleton_bound}) is saturated by the parameters $[[5,1,5;4]]$.
Because the minimum distance of a standard $[[9,1]]$ quantum stabilizer code is at most $3$, this $[[5,1,5;4]]$ code is not equivalent to any standard $9$-qubit code.
\eep
In Example \ref{example:5_1_5_4}, we constructed a $[[5,1,5;4]]$ EAQEC code from a $[[5,1,1]]$ 5-qubit bit flip code,
which is a quantum version of the $[5,1,5]$ classical repetition code.
The result can be generalized to the construction of an $[[n,1,n;n-1]]$ EAQEC code for  $n$ odd from an $[n,1,n]$  classical repetition code as follows.
\bt \label{thm:new code}
    There are $[[n,1,n;n-1]]$ EAQEC codes for  $n$ odd that achieve the quantum singleton bound (\ref{eq:singleton_bound}) and
    are not equivalent to any standard quantum stabilizer code.
\et
\begin{proof}
Suppose $\hat{H}_n$ is an $(n-1)\times n$ parity-check matrix of a classical $[n,1,n]$ repetition code:  
\[
\hat{H}_n= \begin{bmatrix}
1&1&0&\cdots &\cdots &0\\
0&1&1&\ddots &\cdots &0\\
\vdots&\ddots &\ddots&\ddots&\ddots&\vdots\\
0&\cdots &\cdots & 1&1&0\\
0&\cdots &\cdots & 0&1&1
\end{bmatrix}.
\]
Then the $[[n,1,1 ]]$ $n-$qubit bit-flip code has a check matrix
\[
\left[
\begin{array}{c|c}
O& \hat{H}_n\\
\end{array}
\right].
\]
We want to introduce $(n-1)$ simplified generators  to the generating set of the stabilizer group
such that the minimum distance of the code is increased to $n$.
Consider a simplified check matrix
\begin{align*}
H'=\begin{bmatrix}
O & \hat{H}_n\\
\hat{H}_n & O\\
\end{bmatrix},
\end{align*}
By (\ref{eq:number of ebits}), the number of symplectic pairs in $H'$ is
\[
\frac{1}{2}\mbox{rank}(H'\Lambda H'^T)=\mbox{rank}(\hat{H}_n\hat{H}_n^T)=n-1,
\]
for $n$ odd.
It can be verified that $H'$ is a simplified check matrix with minimum distance $n$.
Therefore, there exists a set of symplectic partners of the generators of the stabilizer group of the $n-$qubit bit flip code
such that the minimum distance of the code is $n$.
It is easy to verify that (\ref{eq:singleton_bound}) is saturated by the parameters $[[n,1,n;n-1]]$.
These $[[n,1,n;n-1]]$ codes  are not equivalent to any standard quantum stabilizer code, for there are no standard $[[2n-1,1,n]]$ quantum codes.
\end{proof}

\be
By Theorem \ref{thm:new code} with $n=5$ and
\[
\hat{H}_5= \begin{bmatrix}
1&1&0&0 &0\\
0&1&1&0 &0\\
0&0 & 1&1&0\\
0&0 & 0&1&1
\end{bmatrix},
\]
which is a parity-check matrix of a classical $[5,1,5]$ repetition code, a set of generators of the stabilizer of the $[[5,1,1]]$
5-qubit bit flip code is
\begin{align*}
g_1'=&ZZIII\\
g_2'=&IZZII\\
g_3'=&IIZZI\\
g_4'=&IIIZZ.
\end{align*}
Then we add the following set of generators:
\begin{equation*}
\begin{split}
\tilde{h_1}=&XXIII\\
\tilde{h_2}=&IXXII\\
\tilde{h_3}=&IIXXI\\
\tilde{h_4}=&IIIXX. \label{eq:high_distance_matrix}
\end{split}
\end{equation*}
These generators do not form symplectic pairs.
Observe that $[\tilde{g_i},\tilde{g_j}]=0$, $[\tilde{h_i},\tilde{h_j}]=0$, and $[\tilde{g_i},\tilde{h_j}]=0$
except $\{\tilde{g_1}, \tilde{h_2}\}=0$, $\{\tilde{g_2}, \tilde{h_1}\}=0$, $\{\tilde{g_2}, \tilde{h_3}\}=0$,
$\{\tilde{g_3}, \tilde{h_2}\}=0$, $\{\tilde{g_3}, \tilde{h_4}\}=0$, $\{\tilde{g_4}, \tilde{h_3}\}=0$.
Let $h_1'=\tilde{h_2}\tilde{h_4}=IXXXX$, $h_2'=\tilde{h_1}$, $h_3'=\tilde{h_4}$, and
$h_4'=\tilde{h_1}\tilde{h_3}=XXXXI$.
Then we have
\begin{align*}
&[{g}_i',{g}_j']=0, \mbox{ for all $i, j$},\\
&[{f}_i',{f}_j']=0, \mbox{ for all $i, j$},\\
&\{{g}_i',{f}_i'\}=0, \mbox{ for all $i$},\\
&[{g}_i',{f}_j']=0, \mbox{ for $i\neq j$}.
\end{align*}
Hence we have introduced $4$ simplified generators such that there are $4$ symplectic pairs.
Observe that if  any one of the simplified generators $h_1', h_2', h_3', h_4'$ is removed and $c=3$, the minimum distance instantly drops to $2$.
If two simplified generators $h_1', h_2'$ or $h_1', h_4'$ or $h_2',h_3'$ are removed and $c=2$, the minimum distance further decreases to $1$.
\eep

According to \cite{BDM06}, given a parity-check matrix $\hat{H}$
of an $[n,k,d]$ classical binary  linear code, an
$[[n,2k+c-n,d;c]]$ EAQEC code can be constructed from a simplified
check matrix $H'$, defined as
\begin{align} \label{eq:construction Brun}
H'=\begin{bmatrix}
O & \hat{H}\\
\hat{H} & O\\
\end{bmatrix},
\end{align}
 where the number of ebits $c$ required for this EAQEC code is given by (\ref{eq:number of ebits}).
The family of EAQEC codes in Theorem \ref{thm:new code} can also be obtained by this construction.
%
When $c=n-k,$ the quantum singleton bound
(\ref{eq:singleton_bound}) becomes
\[
n-k\geq d-1,
\]
which is exactly the same as the classical singleton bound.
However, there are no nontrivial classical binary codes achieve the singleton bound from \cite{MS77}.

\subsection{Unitary Row Operators} \label{Sec:Row_operation}
 Since we have the freedom to choose  among
different sets of generators of a stabilizer group, and also the
freedom to choose which ancilla qubits are
replaced by ebits when $c<r$,
we will show that the minimum distance can be
further optimized over these two factors  when $c<r$.
We first discuss the effect of some ``unitary row operators," which preserve the overall commutation relations
(\ref{def:symplectic pairs 1})-(\ref{def:symplectic pairs 4}).

Consider a unitary operator $U=\frac{1}{\sqrt{2}}\left(I+iQ \right)$,
 where $Q$ is a  Pauli operator with eigenvalues $\pm 1$.
It is easy to verify that
\[
 UgU^{\dag}=\left\{
\begin{array}{ll}
    g, & \hbox{ if $[Q,g]=0$;} \\
  iQg, & \hbox{if $\{Q,g \}=0$.} \\
\end{array}%
\right.
\]
We define
$V_{1,2}=V_3V_2V_1$, where
\[V_1= \frac{1}{\sqrt{2}}\left(I+ig_1'h_2' \right), \ V_2= \frac{1}{\sqrt{2}}\left(I-if'_2 \right) ,\]
and
\[V_3= \frac{1}{\sqrt{2}}\left(I-ig'_1 \right) .\]
Then
\[
V_{1,2}g_j' V_{1,2}^{\dag}=\left\{
\begin{array}{ll}
    g_1'g_2', & \hbox{if $j=2$;} \\
    g_j', & \hbox{if $j\neq 2$.} \\
\end{array}%
\right.
\]
Therefore, $V_{1,2}$ is a unitary operator that performs
multiplication of $g_1'$ to $g_2'$, which corresponds to adding
the first row to the second in the simplified check matrix.
On the other hand,
\[
V_{1,2}h_j' V_{1,2}^{\dag}=\left\{
\begin{array}{ll}
    h_2'h_1', & \hbox{if $j=1$;}  \\
    h_j', & \hbox{if $j\neq 1$.} \\
\end{array}%
\right.
\]
Hence a row operation performed on $\{g_1', \cdots, g_r'\}$ induces a row operation performed on $\{h_1', \cdots, h_r'\}$
in order to preserve the commutation relations (\ref{def:symplectic pairs 1})-(\ref{def:symplectic pairs 4}).
We call $V_{1,2}$ a \emph{unitary row operator}.
Later we will need unitary row operators that perform the multiplication of $g_i'$ to $h_j'$, the multiplication
of $\bar{Z}_i'$ to $h_j'$, and the multiplication of $\bar{X}_i'$ to $h_j'$, respectively.
These four types of unitary row operators are summarized in Table \ref{tb:unitary row operation}.
\begin{table}[bht]
  \centering
  \caption{Four types of unitary row operators} \label{tb:unitary row operation}
  \begin{tabular}{|c|c|}
\hline
Type $1.$ &
$Vh_j' V^{\dag}=\left\{
\begin{array}{ll}
    h_l'h_m', & \hbox{if $j=l$;}  \\
    h_j', & \hbox{if $j\neq l$.} \\
\end{array}%
\right.$$Vg_j' V^{\dag}=\left\{
\begin{array}{ll}
    g_m'g_l', & \hbox{if $j=m$;} \\
    g_j', & \hbox{if $j\neq m$.} \\
\end{array}%
\right.$\\
\hline Type $2.$ & $Vh_j' V^{\dag}=\left\{
\begin{array}{ll}
    h_l'g_m', & \hbox{if $j=l$;}  \\
    h_j', & \hbox{if $j\neq l$.} \\
\end{array}%
\right.$$Vh_j' V^{\dag}=\left\{
\begin{array}{ll}
    h_m'g_l', & \hbox{if $j=m$;} \\
    h_j', & \hbox{if $j\neq m$.} \\
\end{array}%
\right.$\\

\hline
Type $3.$ &
    $V h_j' V^{\dag}=\left\{
\begin{array}{ll}
    h_l'\bar{Z}_m', & \hbox{if $j=l$;} \\
    h_j', & \hbox{if $j\neq l$.} \\
\end{array}%
\right.$ $V \bar{X}_j' V^{\dag}=\left\{
\begin{array}{ll}
    g_l'\bar{X}_m', & \hbox{if $j=m$;} \\
    X_j', & \hbox{if $j\neq m$.} \\
\end{array}%
\right.$ \\
\hline
Type $4.$ &
     $V h_j' V^{\dag}=\left\{
\begin{array}{ll}
    h_l'\bar{X}_m', & \hbox{if $j=l$;} \\
    h_j', & \hbox{if $j\neq l$.} \\
\end{array}%
\right.$ $
V \bar{Z}_j' V^{\dag}=\left\{
\begin{array}{ll}
    g_l'\bar{Z}_m', & \hbox{if $j=m$;} \\
    Z_j', & \hbox{if $j\neq m$.} \\
\end{array}%
\right.$\\
\hline
 \end{tabular}
\end{table}
%

When a different set of generators of the stabilizer group is chosen
instead of $\{g_1', \cdots, g_r'\}$,
this is equivalent to performing a unitary row operator $V$, which comprises a series of
unitary row operators of type $1$ on $\{g_1', \cdots, g_r'\}$.
The operation of $V$ on the simplified check matrix $H'$ corresponding to $\{g_1', \cdots, g_r',h_1', \cdots, h_r'\}$
is to multiply $H'$ by a $(2n-2k)\times (2n-2k)$ matrix
of the form
\begin{align*} 
M=\left[
\begin{array}{c|c}
M_Z& O\\
O& M_X\\
\end{array}
 \right].
\end{align*}
If \[M_X= R_mR_{m-1}\cdots R_{1},\] where $R_i'$s are elementary
row operations, then
\[M_Z=R_m^TR_{m-1}^T\cdots R_{1}^T.\]
It can be checked that $MH'$ satisfies (\ref{eq:commutation_relations_H}).
If a set $T=\{t_1,\cdots, t_c\}$ of  $c<r$ ancilla qubits are replaced by ebits,
it is possible that after the operation of $V$, the group $\mathcal{S'_I}=\langle  g_j:j\notin T\rangle$
 changes,  and so does the set $\mathcal{N(S')}-\mathcal{S'_I}$.
In addition, the span of a {subset} of $\{h_1', \cdots, h_r'\}$ can change after the operation of $V$,
though the span of the full set remains unchanged.
This means that if we add less than the maximum amount of entanglement to a code,
we must optimize over such unitary row operations.
Since the group $\mathcal{S'_I}$  and  the set $\mathcal{N(S')}-\mathcal{S'_I}$ remain the same
under the operation of type 1 unitary row operators that operate  on the $h_j'$ for $j\notin T$,
it suffices to assume that the operation $V$ consists of  type 1 unitary row operators that operate only on the $h_j'$ for $j\in T$.


Let $M_V$ be a $c\times r$ matrix such that the $i-$th row of $M_V$ is the $t_i-$th row of $M_Z$ for $i=1,\cdots,c$.
It is obvious that some $M_V$'s have the same effect on the row space of $H'$.
For example, if $c=2,$ $\{g_1'g_2',g_2',\cdots,g_r', h_1', h_1'h_2'\}$ and $\{g_1',g_2',\cdots,g_r', h_1', h_2'\}$
are two different sets of generators
but they generate the same space and hence their corresponding
EAQEC codes have the same minimum distance.
Therefore, a distinct unitary row operation $V$ is assumed to be be represented by a matrix $M_V$ in reduced row echelon form.

\bt \label{thm:unitary row operation}
The operation of $V$ is equivalent to applying a series of type 1 unitary row operators on the $h_j'$ for $j\in T$.
In addition, there are
\begin{align*} \label{eq:number of row operations}
N(r,c)\triangleq\sum_{l_c=0}^{r-c}
\sum_{l_{c-1}=0}^{l_c}\sum_{l_{c-2}}^{l_{c-1}}\cdots
\sum_{l_1=0}^{l_2}  2^{ c(r-c)-\sum_{i=1}^c l_i}
\end{align*}
distinct unitary row operations.
\et
\begin{proof}
The total number of distinct unitary row operations $N(r,c)$ is determined as follows.
If we begin with matrices of the form
\[
\begin{bmatrix}
1&0& \cdots & 0 & \square &\cdots &\square\\
0&1&\cdots & 0 &\square&\cdots &\square\\
\vdots &\vdots &  \ddots& \vdots & \vdots & \ddots &\vdots\\
0&0&\cdots & 1 &\square &\cdots &\square
\end{bmatrix},
\]
where $\square$ can be $0$ or $1$,
there are $2^{c(r-c)}$ distinct unitary row operations.
Now we consider matrices in which the leading ones are shifted to the right.
Let $l_j$ denote the shift amount of the leading 1 of $j$-th row from its initial position for $j=1,\cdots,c$.
It can be observed that $l_j\leq l_i$ if $ j<i$.
For a set $\{l_1,l_2,\cdots,l_c\}$, the number of $\square$ is $c(r-c)-\sum_{i=1}^c l_i$, and
hence there are
$  2^{ c(r-c)-\sum_{i=1}^c l_i}$  distinct unitary row operations.
Therefore, summing over all possible sets of $\{i_1,\cdots, i_c\}$ shows that there is a total of
\begin{align*} \label{eq:number of row operations}
N(r,c)=\sum_{l_c=0}^{r-c}
\sum_{l_{c-1}=0}^{l_c}\sum_{l_{c-2}}^{l_{c-1}}\cdots
\sum_{l_1=0}^{l_2}  2^{ c(r-c)-\sum_{i=1}^c l_i}
\end{align*}
distinct unitary row operations up to Gaussian elimination.
\end{proof}
\be For $r=4$ and  $c=2$, possible matrices of $M_V$ are
\begin{align*}
\begin{bmatrix}
1&0& \square & \square\\
0&1& \square & \square
\end{bmatrix},
\begin{bmatrix}
1&\square& 0 & \square\\
0&0& 1 & \square
\end{bmatrix},
\begin{bmatrix}
1&\square& \square & 0\\
0&0& 0 & 1
\end{bmatrix},
\begin{bmatrix}
0&1& 0 & \square\\
0&0& 1 & \square
\end{bmatrix},
\begin{bmatrix}
0&1& \square & 0\\
0&0& 0 & 1
\end{bmatrix},
\begin{bmatrix}
0&0& 1 & 0\\
0&0& 0 & 1
\end{bmatrix}.
\end{align*}
There are \[2^4+2^3+2^2+2^2+2^1+2^0 = 35\]
distinct row operations.
\eep
The function $N(r,c)$ has a symmetric property as in the following lemma and some closed forms of $N(r,c)$ are listed in Table \ref{tb:closed form}.
\bl
$N(r,c)=N(r,r-c)$ for any $r$ and $0\leq c\leq r.$
\el
\begin{proof}
We prove this lemma by mathematical induction.
Assume $N(l,c)=N(l,l-c)$ holds for $l=1,\cdots,r$ and $c=0,\cdots,l$.
It suffices to assume $c<r-c$ or $2c<r$.
Consider the following $c\times (r+1-c)$ matrix:
\begin{align*}
\begin{array}{c|c}
\underbrace{\begin{array}{ccccc}
1&0& \cdots & 0 & \square\\
0&1&\cdots & 0 & \square\\
\vdots &\vdots &  \ddots& \vdots &\vdots\\
0&0&\cdots & 1 & \square
\end{array}}_{c+1}& \underbrace{\left.\begin{array}{ccc}
 \square &\cdots &\square\\
\square&\cdots &\square\\
 \vdots & \ddots &\vdots\\
\square &\cdots &\square
\end{array}\right.}_{r-c}\\
\end{array}&  \
\end{align*}
Dividing the matrix into two parts, we find that \[N(r+1,c)=N(c+1,c)N(r,c).\]
Now consider the following matrix:
\begin{align*}
\begin{array}{cc}
  \left.\begin{array}{ccccc}
1&0& \cdots & 0 & 0\\
0&1&\cdots & 0 & 0\\
\vdots &\vdots &  \ddots& \vdots &\vdots\\
0&0&\cdots & 1 & 0\\
\end{array}\right.& \left.\begin{array}{ccc}
 \square &\cdots &\square\\
\square&\cdots &\square\\
 \vdots & \ddots &\vdots\\
\square &\cdots &\square
\end{array}\right.\\
\hline
\underbrace{\left.\begin{array}{ccccc}
0&0&\cdots & 0 & 1\\
\end{array}\right.}_{r+1-c}& \underbrace{\left.\begin{array}{ccc}
\square &\cdots &\square
\end{array}\right.}_{c}\\
\end{array} & \
\end{align*}
Similarly, we have \[N(r+1,r+1-c)=N(c+1,1)N(r,r-c).\]
From the above two equations, we have
$N(r+1,c)=N(r+1,r+1-c)$.
By induction, we obtain the lemma.
\end{proof}

%
\begin{table}[bht]
  \centering
  \caption{Closed forms of $N(r,c)$ for different $c$} \label{tb:closed form}
  \begin{tabular}{|c|c|}
\hline
 $c$ & $N(r,c)$\\
\hline
$0$ or $r$ &  $1$ \\
\hline
      $1$ or $r-1$ & $2^{r}-1$\\
\hline
      $2$ or $r-2$& $ \frac{2}{3}4^{r-1}-2^{r-1}+\frac{1}{3}$\\
\hline
$3$ or $r-3$ &  $ \frac{1}{21}8^{r-1}-\frac{2}{3}4^{r-2}+\frac{1}{3}2^{r-2}-\frac{1}{21}$\\
\hline
$4$ or $r-4$ &  $\frac{4}{315}16^{r-2}- \frac{1}{21}8^{r-2}+\frac{2}{9}4^{r-3}-\frac{1}{21}2^{r-3}+\frac{1}{315}$\\
\hline
$5$ or $r-5$ &  $\frac{1}{9765}32^{r-2}-\frac{4}{315}16^{r-3}+ \frac{1}{63}8^{r-3}-\frac{2}{63}4^{r-4}+\frac{1}{315}2^{r-4}-\frac{1}{9765}$\\
\hline
$6$ or $r-6$ &  $\frac{8}{615195}64^{r-3}-\frac{1}{9765}32^{r-3}+\frac{4}{945}16^{r-4}- \frac{1}{441}8^{r-4}+\frac{1}{945}4^{r-5}-\frac{1}{9765}2^{r-5}+\frac{1}{615195}$\\
\hline
 \end{tabular}
\end{table}

On the other hand, it can be observed that the selection operator $W$ in the previous subsection can be
decomposed as a series of unitary row operators of type 2, type 3, and type 4,  for
the matrix $A$ determines a series of type 4 unitary row operators, the matrix $B$   determines a series of type 3 unitary row operators,
and the symmetric matrix $M$, satisfying $C= B A^T +M$, determines a series of type 2 unitary row operators.
Actually, unitary row operators of type 2 do not affect
the set $\mathcal{N(S')}-\mathcal{S'_I}$ or the error-correcting
ability, and the symmetric matrix $M$ can be eliminated.
If a set $T=\{t_1,\cdots, t_c\}$ of  $c<r$ ancilla qubits are replaced by ebits,
it can be verified that $\mathcal{N(S')}=\langle g_j: j\notin T, \bar{Z}_1, \cdots, \bar{Z}_k,  \bar{Z}_1, \cdots, \bar{Z}_k   \rangle$
 remains unchanged
under the operation of type 3 and type 4 unitary row operators  on the $h_j'$ for $j\notin T$.
It suffices to assume that the operation $W$ consists of  type 3 and type 4 unitary row operators that operate only on the $h_j'$ for $j\in T$.
To sum up, we have the following theorem.
\bt \label{thm:selection operator}
The operation of $W$ is equivalent to applying a series of type 4 unitary row operators, followed by a series of type 3
unitary row operators, on the $h_j'$ for $j\in T$.
In addition, there are $2^{2ck}$ distinct selection operators with \[C= B A^T.\] 
\et
\noindent
Combining the effects of the unitary row operation $V$ with the selection operator $W$ in the previous section,
we can optimize an encoding operation of the form $U=VU_EW$  over
\[2^{2ck}N(r,c)\]
possibilities.
We call this the \emph{encoding optimization procedure} for EAQEC codes.  Note that we can find another unitary row operator $W'$ corresponding to $W$ such that $W'U_E$   and $U_EW$ are equivalent encoding operators.  While $W$ operates on the raw stabilizer generators and logcial operators, $W'$ operates on the encoded stabilizer generators and logical operators.  Hence, we can also solve the optimization problem for an operator of the form $U=VW'U_E$ (which is what we actually do in practice, combining $VW'$ into a single optimization).

\subsection{Results of the Encoding Optimization Procedure}
\label{sec:result}
We apply the encoding optimization procedure to a $[[7,1,3]]$ quantum BCH code \cite{gb99,AKS05} and
  Shor's $[[9,1,3]]$ code \cite{Shor95} and the results are shown in Table \ref{tb:7BCHcode} and Table \ref{tb:9Shor},
  where $d_{opt}$ is the minimum distance of the optimized EAQEC codes,
$d_{std}$ is the highest  minimum distance of an  $[[n+c,k]]$ standard stabilizer code, and $N_{opt}$ is the number of encoding operators
that give an EAQEC code with minimum distance $d_{opt}$.
\be
The  check matrix  of a standard $[[7,1,3]]$ quantum BCH code adopted in the encoding optimization procedure  is
\[
\begin{bmatrix}
0000000 & 1001011\\
0000000 & 0101110\\
0000000 & 0010111\\
1001011 & 0000000\\
1100101 & 0000000\\
1011100 & 0000000\\
\end{bmatrix}.
\]
\begin{table}[bht]
  \centering
  \caption{ Optimization over the $[[7,1,3]]$ quantum BCH code} \label{tb:7BCHcode}
  \begin{tabular}{|c|c|c|c|c|}
\hline
      $c$&    $d_{opt}$& $d_{std}$ & $N_{opt}$& $2^{2ck}N(r,c)$  \\
      \hline
6 &  7 & 5 &36  &4096\\
5 &  5 & 5 &31920  & 64512 \\
4 &  5 & 5 &39522  & 166656 \\
3 &  5 & 4 &4332   &89280 \\
2 &  5 & 3 &14  &10416  \\
1 &  3 & 3 &252&  252\\
\hline
 \end{tabular}
\end{table}
As shown in Table \ref{tb:7BCHcode}, the parameters $[[7,1,7;6]]$, $[[7,1,5;3]]$ and
$[[7,1,5;2]]$ achieve the quantum singleton bound for EAQECC
(\ref{eq:singleton_bound}) and are not equivalent to any standard
quantum stabilizer code. We would like to compare these two
EAQEC codes to a
competing EAQEC code with $n=7$ and $d=5$ by the construction of
\cite{BDM06} or the binary version (\ref{eq:construction Brun}).
According to Grassl's table \cite{Grassl}, a classical linear
code over $GF(4)$ (or $GF(2)$) that meets our requirement is a
$[7,2,5]$ linear quaternary code, which can be used to
construct a $[[7,2,5; 5]]$ EAQEC code.
This means that the  $[[7,1,5;2]]$ and  $[[7,1,5;3]]$ EAQEC codes cannot be
obtained by the construction of \cite{BDM06}.
Therefore, the  $[[7,1,5;2]]$ and $[[7,1,5;3]]$ EAQEC codes are new.

In addition, all the $[[7,1,5;2]]$ EAQEC codes we found are degenerate codes,
for some simplified stabilizer generators are of weight $4$ from the check matrix.
For example,  a simplified check matrix and its simplified logical matrix of a $[[7,1,5;2]]$ EAQEC code are
\[
\begin{bmatrix}
0000000 &1001011\\
0000000 &1100101\\
0000000 &0010111\\
1001011 &0000000\\
1100101 &0000000\\
0010111 &0000000\\
\hline
1000011 &0100011\\
1101000 &0010010\\
\end{bmatrix}
,
\begin{bmatrix}
1001011 &0100011\\
1101000 &1001011\\
\end{bmatrix},
\] with $T=\{1,4\}$.
On the other hand, all the $[[7,1,7;6]]$ EAQEC codes are nondegenerate codes,
while the $[[7,1,5;3]]$, $[[7,1,5;4]]$,$[[7,1,5;5]]$  EAQEC codes can be degenerate or nondegenerate.

\eep

\be
The check matrix of Shor's $[[9,1,3]]$ code is
\[\begin{bmatrix}
000000000 &110000000\\
000000000 &011000000\\
000000000 &000110000\\
000000000 &000011000\\
000000000 &000000110\\
000000000 &000000011\\
111111000 &000000000\\
000111111 &000000000\\
\end{bmatrix}.\]
As can be seen in Table  \ref{tb:9Shor}, the parameters
$[[9,1,9;8]]$, $[[9,1,7;5]]$ and $[[9,1,7;4]]$ achieve the quantum
singleton bound for EAQECC (\ref{eq:singleton_bound}) and are not
equivalent to any standard quantum stabilizer code. A competing
EAQEC code with $n=9$ and $d=7$ by the construction of
\cite{BDM06} is a $[[9,1,7;6]]$ EAQEC code, obtained from a
$[9,2,7]$ linear quaternary code in Grassl's table.
Therefore, the
$[[9,1,7;5]]$ and $[[9,1,7;4]]$ EAQEC codes are new.
All the
$[[9,1,5;2]]$, $[[9,1,5;3]]$, $[[9,1,7;4]]$, $[[9,1,7;5]]$ and
$[[9,1,7;6]]$ codes are degenerate codes, and all the $[[9,1,9;8]]$
codes are nondegenerate codes, while the $[[9,1,7;7]]$ codes can
be degenerate or nondegenerate.

\begin{table}[bht]
  \centering
  \caption{ Optimization over  Shor's $[[9,1,3]]$  code} \label{tb:9Shor}
  \begin{tabular}{|c|c|c|c|c|}
\hline
      $c$&    $d_{opt}$& $d_{std}$ & $N_{opt}$& $2^{2ck}N(r,c)$  \\
      \hline
8  & 9 &  7 &    256& 65536 \\
7  & 7 &  6 &    330624 & $4.17\times10^6$\\
6  & 7 &  6 &    278904 & $4.42\times 10^7$ \\
5  & 7 &  6 &    17748  & $9.94\times 10^7$ \\
4  & 7 &  5 &    132 & $5.14\times 10^7$   \\
3  & 5 &  5 &    69777 &  $6.21\times 10^6$ \\
2  & 5 &  5 &    201 & $1.72\times 10^5$   \\
\hline
 \end{tabular}
\end{table}


\eep

\section{A Random Search Algorithm for Encoding Optimization}
\label{sec:random_search}
It is easy to check that
\begin{align*}
2^{c(r-c)}\leq N(r,c)\leq {r\choose c} 2^{c(r-c)}.
\end{align*}
Hence
\begin{align*}
2^{c(n+k-c)}\leq 2^{2ck}N(r,c)\leq {r\choose c} 2^{c(n+k-c)}.
\end{align*}
A complete encoding optimization procedure for a $[[n,k,d]]$
standard code becomes impossible when $n+k$ becomes large. Hence
we consider a random search algorithm for the encoding
optimization procedure in this section.


We define the \emph{weight enumerator}  of the set $N(\mathcal{S}')-\mathcal{S}'_I$ as $f(x)=\sum_{i=0}^n a_i x^i$, where
$a_i$ is the number of elements of weight $i$ in  $N(\mathcal{S}')-\mathcal{S}'_I$.
Suppose $H'$ is a simplified check matrix of the EAQEC code corresponding to $\mathcal{S}'$.
Then we  define a \emph{merit function} $m: H'\mapsto \mathbb{R}$ by
\[m(H')= \sum_{i=1}^b a_i,\]
where $b=\lfloor \frac{n+c-k-2}{2} \rfloor$ is an upper bound on the minimum distance for the given parameters $n,k,c$.
This merit function was chosen to try to maximize the number of correctable errors in the set of likely errors for a typical memoryless channel;
but of course, other merit functions could be chosen.
The random search algorithm for encoding optimization is as follows:
\begin{enumerate}[1)]
    \item Apply Wilde's encoding circuit algorithm  to a given check matrix $H_0$ to obtain the encoding operator $U_E$.
    \item Apply $U_E$ to the simplified check matrix (\ref{eq:initial_stabilizer_check_matrix_EAQECC})
    and the simplified logical matrix (\ref{eq:initial_logical_matrix_EAQECC}), and denote the new matrices by $H_0'$ and $L_0'$, respectively.
    \item Compute the merit function $m(H_0')$.
    \item Randomly choose a set $T=\{t_1,\cdots, t_c\}$.
    \item Randomly generate the matrices $A$ and $B$ according to Theorem \ref{thm:selection operator}. 
    \item Apply a selection operator based on $A$ and $B$, and obtain a new simplified logical matrix $L'$.
    \item Randomly generate a matrix $M_V$ according to Theorem \ref{thm:unitary row operation}.
    \item Apply the unitary row operator defined by $M_V$ to obtain a new simplified logical matrix $L'$.
    \item Compute the minimum distance $d$ of the EAQEC code defined by $H'$ and $L'$.
    \item Compute the merit function $m(H')$.
    \item If $m(H')<m(H_0')$, set $H_0'\longleftarrow H'$.
    \item If $d=b=\lfloor \frac{n+c-k-2}{2} \rfloor$, or the search has repeated more than a maximum number of times, stop.
    \item  Else, Go to step 4).
\end{enumerate}
Since the minimum distance might not be the best  measure of  the error-correcting ability of a quantum error-correcting code,
a different merit function can be adopted.
Here we simply chose the merit function $m()$ to encourage EAQEC codes with fewer low-weight elements in $N(\mathcal{S}')-\mathcal{S}'_I$.
Some examples of this random search algorithm follow:


\be
Applying the random search algorithm to the $[[7,1,3]]$ quantum BCH code and  Shor's $[[9,1,3]]$ code,
we obtain the same results as those in Table  \ref{tb:7BCHcode} and
 Table  \ref{tb:9Shor} by the complete encoding optimization procedure.
We list in Table  \ref{tb:7BCHcode random} and
 Table  \ref{tb:9Shor random} the average number of loops, $N_{avg}$,
 to obtain these parameters (which we estimate by performing the algorithm $10^4$ times  for each $c$).
 Note that $N_{avg}$ is close to $2^{2ck}N(r,c)/N_{opt}$ for the $[[7,1,3]]$ quantum BCH code,
 except for the case $c=2$, where $N_{avg}$ is about half of  $2^{2ck}N(r,c)/N_{opt}$.
For Shor's $[[9,1,3]]$  code, $N_{avg}$ is much less than  $2^{2ck}N(r,c)/N_{opt}$ in most cases.

\begin{table}[bht]
  \centering
  \caption{ Average number of trials of the random search algorithm on the $[[7,1,3]]$ quantum BCH code} \label{tb:7BCHcode random}
  \begin{tabular}{|c|c|c|c|}
\hline
      $c$&   $d_{opt}$&  $2^{2ck}N(r,c)/N_{opt}$& $N_{avg}$\\
      \hline
6 &  7&    113.78& 109.37 \\
5 &  5& 2.02 &  2.02\\
4 &  5& 4.22& 4.13\\
3 & 5& 20.61 &  18.63\\
2 &  5& 744 &392.57 \\
\hline
 \end{tabular}
\end{table}
\begin{table}[bht]
  \centering
  \caption{  Average number of trials of the random search algorithm on  Shor's $[[9,1,3]]$  code} \label{tb:9Shor random}
  \begin{tabular}{|c|c|c|c|}
\hline
      $c$& $d_{opt}$&   $2^{2ck}N(r,c)/N_{opt}$ &  $N_{avg}$\\
      \hline
8  &  9&  256 &259.97\\
7  &  7&  12.61& 12.47\\
6  &  7& 158.48& 119.25 \\
5  &  7& 5600.63& 1188.08 \\
4  &  7& 389393.94& 7577.95\\
3  &  5&   88.99& 18.48\\
2  &  5& 855.72&  52.46 \\
\hline
 \end{tabular}
\end{table}
\eep

\be We applied the random search algorithm  to a
standard $[[15,7,3]]$ quantum BCH code and the results are shown in Table \ref{tb:15BCHcode}.
\begin{table}[bht]
  \centering
  \caption{Optimization over a $[[15,7,3]]$ Quantum BCH code} \label{tb:15BCHcode}
  \begin{tabular}{|c|c|c|c|}
\hline
      $c$&    $d_{opt}$& $d_{std}$ & $2^{2ck}N(r,c)$\\
      \hline
8&   6&   5-6  & $5.19\times 10^{33}$\\
7&   5&   5-6& $8.08\times 10^{31}$\\
6&   5&   5-6& $2.08\times 10^{29}$\\
5&   4&  4-5& $1.14\times 10^{26}$\\
4&   4&   4-5& $1.44\times 10^{22}$\\
7&   3&   4& $4.27\times 10^{17}$\\
\hline
 \end{tabular}
\end{table}
Note that we did not find the fully optimized parameters in this case, since the
complexity is very high.
However, compared with the $[[15,3,5;4]]$ EAQEC
code, obtained by the construction of (\ref{eq:construction Brun}) from a
$[15,7,5]$ classical BCH code, the $[[15,7,5;7]]$ and the
$[[15,7,5;6]]$ EAQEC codes have $4$ more information qubits
at the cost of $3$ and $2$ more ebits, respectively. The
$[[15,7,6;8]]$ EAQEC code has $4$ more information qubits and a
higher minimum distance at the cost of $4$ more ebits.
In addition, the $[[15,7,6;8]]$ EAQEC code is not equivalent to any known standard quantum stabilizer code.

On the other hand, the classical linear quaternary
$[15,9,5]$ code and $[15,8,6]$ code in  Grassl's table
 can be used to construct a $[[15,9,5; 6]]$ EAQEC
code and a $[[15,9,6; 8]]$ EAQEC code  by the construction of
\cite{BDM06}, respectively. These two codes are better than the
$[[15,7,6;8]]$  EAQEC code we obtained.
This may be because these codes were not fully optimized,  but
BCH codes in any case need not give the best possible EAQEC codes, even using the
encoding optimization procedure.

\eep

\be
We applied the random search algorithm  to Gottesman's $[[8,3,3]]$ code \cite{Got96} and the $[[13,1,5]]$ quantum QR
code \cite{CRSS97,LLu07}, and the results
are shown in Table \ref{tb:8Got}  and Table \ref{tb:13QR}, respectively.
By the construction of \cite{BDM06}, the $[8,3,5]$, $[13,3,9]$, $[13,4,8]$, $[13,5,7]$
classical linear quaternary codes in Grassl's Table can be transformed to
$[[8,2,5;4]]$, $[[13,3,9;10]]$, $[[13,0,8;5]$, $[[13,1,7;4]]$ EAQEC codes, respectively.
Hence  the $[[8,3,5;5]]$,  $[[13,1,11;11]]$,  $[[13,1,11;10]]$,  $[[13,1,9;9]]$,  $[[13,1,9;8]]$  EAQEC
codes are new, and are not equivalent to any standard quantum stabilizer code.
\begin{table}[bht]
  \centering
  \caption{Optimization over Gottesman's $[[8,3,3]]$  code} \label{tb:8Got}
  \begin{tabular}{|c|c|c|c|c|c|c|c|}
\hline
      $c$&    $d_{opt}$& $d_{std}$ & $2^{2ck}N(r,c)$\\
      \hline
5  & 5 &  4   &    $1.07\times 10^{9}$\\
4  & 4 &  4   &      $5.20\times 10^{7}$\\
3  & 4 &  3   &     $4.06\times 10^{7}$\\
2  & 3 &  3   &    $6.34\times 10^{6}$ \\
\hline
 \end{tabular}
\end{table}
\begin{table}[bht]
  \centering
  \caption{Optimization over the $[[13,1,5]]$  quantum QR code} \label{tb:13QR}
  \begin{tabular}{|c|c|c|c|c|c|c|c|}
\hline
      $c$&    $d_{opt}$& $d_{std}$ & $2^{2ck}N(r,c)$\\
      \hline
12&  13&  9&     $1.68\times 10^{7}$\\
11&  11&  8-9&   $1.71\times 10^{10}$\\
10&  11&  7-9&   $2.92\times 10^{12}$\\
9 &  9&   7-8&   $1.07\times 10^{14}$\\
8 &  9&   7&     $9.12\times 10^{14}$\\
7 &  7&   7&     $1.87\times 10^{15}$\\
6 &  7&   7&     $9.44\times 10^{14}$\\
5 &  7&   7&     $1.17\times 10^{14}$\\
4 &  7&   7&     $3.56\times 10^{12}$\\
\hline
 \end{tabular}
\end{table}
\eep


\section{Circulant Construction of EAQEC codes}
\label{sec:circulant_construction}
The construction in Theorem \ref{thm:new code} and the construction (\ref{eq:construction Brun}) resemble the CSS construction \cite{CS96,Ste96},
 while standard quantum stabilizer codes with a high minimum distance are usually non-CSS codes.
In this section we give a different construction of EAQEC codes,
with a simplified check matrix that differs from the simplified
check matrix in construction (\ref{eq:construction Brun}).
We construct the simplified check matrix directly,
rather than starting from a classical binary code.

Let $H'$ be a $r\times 2n$ simplified check matrix cyclicly generated by a binary $2n-$tuple $\textbf{a}=a_0 a_1 \cdots a_{2n-2}a_{2n-1}$:
\begin{equation*}
 H'=[ H_X' | H_Z']=
\left[ \begin{array}{cccc|cccc}
 a_0& a_1 & \ldots & a_{n-1}&a_n& a_{n+1} & \ldots & a_{2n-1}\\
 a_{1}& a_2 & \ldots & a_{n}& a_{n+1}& a_{n+2} & \ldots & a_{0}\\
 \vdots & \vdots & \ddots & \vdots &\vdots & \vdots & \ddots & \vdots \\
 a_{r-1}& a_r & \ldots & a_{r+n-2} &a_{r+n-1}& a_{r+n} & \ldots & a_{r-2}\\
 \end{array}\right].
\end{equation*}
If the rank of $H'$ is exactly $r$, then $c=\frac{1}{2}\mbox{rank}(H'\Lambda H')$
and $H'$ defines an $[[n, n+c-r,d;c ]]$ EAQEC code for some minimum distance $d$.
For example, a $[[6,1,4;1]]$ code is
constructed by $\textbf{a}=001110101110$ with the simplified check matrix
\[
\begin{bmatrix}
001110&101110\\
000111&010111\\
100011&101011\\
110001&110101\\
111000&111010\\
011100&011101\\
\end{bmatrix}.
\]
We call this the \emph{circulant} construction of EAQEC codes, which is
used for standard stabilizer codes in \cite{LLu07}.

We examined the simplified check matrices cyclicly generated by every possible binary $2n-$tuple $\textbf{a}$ by computer for $n=4, \cdots, 10$ and $r\leq 2(n-1)$.
Parameters of EAQEC codes not equivalent to any standard quantum stabilizer codes are listed in Table \ref{tb:eaqecc_circulant}.
The parameters $[[4,0,4;2]]$, $[[ 4, 1, 3; 1]]$, $[[5,0,4;2]]$, $[[5,1,4;3]]$, $[[ 5, 1, 5; 4]]$, $[[6,0,6;4]]$, $[[ 6, 2, 3; 1]]$, $[[6,2,4;3]]$, $[[ 6, 0, 4; 1]]$, $[[ 6, 1, 5; 4]]$, $[[ 7, 4, 3; 2]] $, $[[7,1,6;5]]$,
$[[7, 1, 7;6]]$, $[[ 8, 5, 3; 2]]$, $[[9, 1, 9;8]]$, and $[[10,0,10;8]]$
also saturate the quantum singleton bound (\ref{eq:singleton_bound}).

\begin{table}[bht]
\caption{
Parameters of $[[n,k,d;c]]$ EAQEC codes not equivalent to any standard $[[n+c,k]]$ codes.}
\[
         \begin{tabular}{|c|c|}
\hline
        $n$ &  $[[n, k, d;c]]$  \\
\hline
        $4$ & $[[4,0,4;2]]$, $[[ 4, 1, 3; 1]]$\\
\hline
        $5$ &  $[[ 5, 1, 5; 4]]$, $[[5, 1, 4;3]]$, $[[5,1,4;2]]$, $[[5, 0, 4;2]]$, $[[5,2,3;2]]$ \\
\hline
        $6$ & $[[6,0,6;4]]$, $[[6, 1, 5;4]]$, $[[6, 1, 4;3]]$, $[[6, 2, 4;3]]$,
        $[[6, 0, 4;1]]$, $[[6, 2, 3;1]]$      \\
\hline
        $7$ & $[[7, 1, 7;6]]$, $[[7, 2, 5;5]]$, $[[7, 0, 6;4]]$, $[[7, 3, 4;4]]$,$[[7, 1, 4;2]]$, $[[7, 3, 4;3]]$, $[[ 7, 4, 3; 2]]$       \\
\hline
        $8$ & $[[8, 0, 8;6]]$, $[[8, 1, 6;6]]$, $[[8, 0, 6;5]]$, $[[8, 2, 6;6]]$, $[[8, 1, 6;5]]$, $[[8, 0, 6;4]]$,\\
        &  $[[8, 3, 5;5]]$, $[[8, 2, 5;4]]$, $[[8, 1, 4;1]]$, $[[8, 3, 4;3]]$,$[[ 8, 5, 3; 2]]$,\\
\hline
        $9$ & $[[9, 1, 9;8]]$, $[[9, 0, 7;6]]$, $[[9, 1, 7;6]]$, $[[9, 1, 7;7]]$, $[[9, 2, 6;6]]$, $[[9, 1, 6;5]]$, \\
        &  $[[9, 0, 6;4]]$, $[[9,1,6;6]]$, $[[ 9, 2, 5; 4]]$, $[[ 9, 5, 3; 1]]$,\\
\hline
        $10$ &  $[[10,0,10;8]]$, $[[10, 1, 8;8]]$, $[[10, 0, 8;7]]$,  $[[10,0,8;6]]$, $[[ 10, 0, 7; 5]]$, $[[10, 1, 7;6]]$, \\
        & $[[10, 2, 7;7]]$, $[[10, 1, 6;5]]$, $[[10, 3, 6;7]]$, $[[10, 0, 6;3]]$, $[[10, 3, 6;6]]$, $[[10, 2, 6;5]]$, \\
        &  $[[10, 1, 6;4]]$, $[[10, 4, 5;5]]$, $[[10,2,5;2]]$, $[[10, 4, 5;4]]$,  $[[10, 2, 5;3]]$,\\
\hline
        \end{tabular}
\]
 \label{tb:eaqecc_circulant}
\end{table}


\section{Discussion}

This paper has studied how entanglement can be used to increase
the minimum distance of quantum error-correcting codes. We
demonstrated the encoding optimization procedure for EAQEC codes
obtained by adding ebits to standard quantum stabilizer codes.
The four types of unitary row operators play an important role in
this encoding optimization procedure, and also help to clarify the
properties of EAQEC codes and their relationship to standard codes.
 Some applications of the encoding
optimization procedure were found to have promising results:
we found
$[[7,1,5;2]]$ and $[[7,1,5;3]]$ EAQEC codes from quantum BCH codes,
and $[[9,1,7;4]]$ and $[[9,1,7;5]]$ EAQEC codes from Shor's $9-$qubit
code, together with a family of $[[n,1,n;n-1]]$ EAQEC codes for  $n$ odd, all of which
achieve the quantum singleton bound.
Several EAQEC codes found by this encoding optimization procedure are
also degenerate codes.
This procedure serves as an EAQEC code
construction method for given parameters $n,k,c$.

The encoding optimization procedure has very high complexity. However, it might be useful to  further investigate it for
specific families of codes that have special algebraic structures, such as  quantum BCH codes and quantum Reed-Muller codes.
For example, if we added ebits to a quantum Reed-Muller code (by the CSS construction with a classical $RM(r,m)$ code \cite{MS77}),
we found that its simplified logical matrix $L'$ is, like a check matrix of a CSS code, of the form
\[
\begin{bmatrix}
O & L_Z'\\
L_X' &O\\
\end{bmatrix}.
\]
By examining the symplectic relations, we found that the classical code, generated by $L_Z'$ or $L_X'$,
together with the $RM(r,m)$ code, generate a subcode of $RM(r',m)$, which contains more lower-weight codewords.
The minimum distance of the EAQEC code might be increased approximately by a factor $\frac{3}{2}$
similar to the method used in \cite{Ste98} or \cite{LLu07}.
From a $[[16,6,4]]$ quantum Reed-Muller code, we obtained a $[[16,6,6;10]]$ EAQEC code by
constructing the matrix $M_V$ such that the above argument could be applied.
However, we obtained a $[[16,6,7;10]]$ EAQEC code by applying the random search algorithm for the encoding optimization procedure.
How to directly construct a matrix $M_V$ that leads to EAQEC codes with high minimum distances is a subject of ongoing research.

When the complexity becomes large, it is almost impossible to optimize over all $2^{2ck}N(r,c)$ encoding operators.
The random search algorithm seems to be the only method to achieve good (but suboptimal) results for EAQEC codes.
For different parameters $n,k,c$, the merit function should be carefully chosen.
The best choice of merit function for a given application is also a subject of future work.
A search algorithm for specific EAQEC codes could be developed.
While the encoding optimization procedure in this paper applies to a standard quantum stabilizer code,
it is possible to construct a similar encoding optimization algorithm for adding ebits to other
EAQEC codes that have ancilla qubits which are not ebits.
Much work remains to be done in finding the best possible EAQEC codes for different applications.




\bibliographystyle{IEEEtran}
\bibliography{IEEEabrv,qecc}

\vfill\eject
\end{document}